\documentclass[preprint,seceq]{jpsj2}

\makeatletter
\@addtoreset{equation}{section}
\makeatother
\def\imagunit{{\rm i}}
\def\napier{{\rm e}}
\def\MC{{\cal M}}

\recdate{May 17, 2005}

\title{%
Max-Plus Algebra for Complex Variables and Its Application to
Discrete Fourier Transformation}
\author{%
Tetsu \textsc{Yajima}\thanks{yajimat@is.utsunomiya-u.ac.jp},
Keisuke \textsc{Nakajima} and
Naruyoshi \textsc{Asano}}
\inst{%
Department of Information Science, Faculty of Engineering, 
Utsunomiya University,\\ Yoto 7-1-2, Utsunomiya, Tochigi 321-8585}
\abst{%
A generalization of the max-plus transformation, which is known as a method
to derive cellular automata from integrable equations, 
is proposed for complex numbers.
Operation rules for this transformation is also studied for general number of
complex variables.
As an application, the max-plus transformation is applied to the
discrete Fourier transformation.
Stretched coordinates are introduced to obtain the max-plus transformation
whose imaginary part coinsides with a phase of the discrete Fourier
transformation.}
\kword{%
Ultradiscretization,
complex variables,
max-plus algebra,
discrete Fourier transformation,
cellular automata,
stretched coordinates}
\begin{document}
\maketitle
\section{Introduction}
The importance of studies on discrete systems generated from integrable
equations has recently attracted considerable attention.
Because these discrete systems conserve good properties of integrable
equations, 
such systems, known as integrable difference equations and soliton 
callular automata\cite{TS}, are important as models to describe engineering
problems.
\par
Ultradiscretization is introduced as one of the systematic methods
to derive discrete systems from differential equations\cite{TTMS,MSTTT}.
It is based on an exponential transformation of field variables $u$
into $U$,
\begin{subequations}\label{eq:sec01-mp-trans-all}
\begin{equation}\label{eq:sec01-mp-trans-01}
u=\napier^{U/\varepsilon},\quad
\mbox{$\varepsilon>0$},
\end{equation}
followed by the limiting process $\varepsilon\to+0$.
By virtue of a relation
\begin{equation}
\lim_{\varepsilon\to+0}\varepsilon\log\left(
\napier^{A/\varepsilon}+\napier^{B/\varepsilon}\right)
=\max(A,B),
\end{equation}
where
\begin{equation}\label{eq:sec01-maxrel-def}
\max(A,B)=\begin{cases}
A&(A\ge B),\\ B&(A<B),\\
\end{cases}
\end{equation}
\end{subequations}
we can transfer summation and product of positive quantities $a$ and $b$.
Let us set $a=e^{A/\varepsilon}$ and $b=e^{B/\varepsilon}$ respectively,
and then consider the logarithms of given equations, 
multiply $\varepsilon$ and take limit of $\varepsilon\to +0$.
Variables $A$ and $B$ may depend on parameter $\varepsilon$,
and if they remain finite at $\varepsilon\to+0$, we have
\begin{subequations}\label{eq:sec02-mp-real-all}
\begin{align}
a+b&=\napier^{A/\varepsilon}+\napier^{B/\varepsilon}\to \max(A,B),
\label{eq:sec02-mp-real-summation}\\
ab&=\napier^{(A+B)/\varepsilon}\to A+B.
\label{eq:sec02-mp-real-product}
\end{align}
\end{subequations}
This procedure is usually called ``max-plus transformation''.
Many nonlinear equations have been transferred to full-discrete systems
by this method, and the relations between them have been studied 
intensively\cite{KT,TTM}.
\par
However, it is not sufficient to consider the max-plus algebra
only for real variables, because all the variables sould be positive
in order to ensure the relevance of the transformation and limit
defined in eqs.\ (\ref{eq:sec01-mp-trans-all}).
Since the physical variables can be negative, this restriction is
fatal from the viewpoint to discretize solutions of difference
equations directly.
\par
In this paper, we aim to introduce a well-defined limiting process
for complex variables, which corresponds to a generalization of
the max-plus transformation for real numbers.
By considering eqs.\ (\ref{eq:sec01-mp-trans-all}) for complex
variables, we can obtain a new max-plus type algebra.
We can discretize the transformation $U$ of the solutions of
difference equations automatically, if the initial values are
properly chosen discrete.
This method eliminates the difficulty of the max-plus transformation
for real variables.
Moreover, we can apply the max-plus transformation to discrete Fourier
transformation (hereafter we shall abbreviate as DFT), 
by which we can solve various differential equations.
\par
This paper is organized as follows.
In the next section, we shall introduce a max-plus algebra for two
complex variables.
A generalization of this algebla for more than two variables is
given in \S3.
In \S4, we apply this algebra to DFT.
The final section is devoted to summary and discussions.
\section{Max-Plus Type Transformation for Two Complex Variables}
Let us consider the generalization of max-plus transformation for the
sum of two complex variables $u_1$ and $u_2$:
\begin{equation}\label{eq:sec02-comp-summation}
u\equiv u_1+u_2.
\end{equation}
Parallel to the method for real variables, we introduce new complex 
quantities $U_1$ and $U_2$ corresponding respectively
\begin{equation}\label{eq:sec02-variable-trans}
\mbox{$u_1=\napier^{U_1/\varepsilon}$ and $u_2=\napier^{U_2/\varepsilon}$,}
\end{equation}
where $\varepsilon$ is a positive number.
Hereafter, we shall write $\varepsilon\log u$ as $W(\varepsilon)$
for simplicity, and are going to derive its explicit expression.
Substituting the relations (\ref{eq:sec02-variable-trans}) into
(\ref{eq:sec02-comp-summation}),
we have
\begin{equation}\label{eq:sec02-mp-trans01}
W(\varepsilon)=
\varepsilon\log\left(\napier^{U_1/\varepsilon}+
\napier^{U_2/\varepsilon}\right).
\end{equation}
By the definition of the logarithms on the complex plane, 
we can write down the real and the imaginary parts of the right-hand
side of eq.\ (\ref{eq:sec02-mp-trans01}) to see
\begin{equation}\label{eq:sec02-mp-trans-rhs}
W(\varepsilon)=
\varepsilon\ln\left|\napier^{U_1/\varepsilon}+\napier^{U_2/\varepsilon}\right|
+\imagunit\varepsilon
\arg\left(\napier^{U_1/\varepsilon}+\napier^{U_2/\varepsilon}\right).
\end{equation}
We express $U_1$ and $U_2$ as $U_j=x_j+\imagunit y_j$ ($j=1,2$), 
where $x_j$, $y_j$ are real numbers, and we assume $x_1>x_2$
for the nonce.
First, we consider the real part of $W(\varepsilon)$.
From direct calculation, we have
\begin{equation}
\begin{split}
\left|\napier^{U_1/\varepsilon}+
\napier^{U_2/\varepsilon}\right|&=
\left[\left(\napier^{2x_1/\varepsilon}+
\napier^{2x_2/\varepsilon}\right)(1+\Delta)\right]^{1/2},\\
\Delta&\equiv{\rm sech}\,\frac{x_1-x_2}{\varepsilon}
\cos\frac{y_1-y_2}{\varepsilon}.
\end{split}
\label{eq:sec02-mp-trans-lnarg}
\end{equation}
From eqs.\ (\ref{eq:sec02-mp-trans-rhs})
and (\ref{eq:sec02-mp-trans-lnarg}), we find the real part of 
$W(\varepsilon)$ as
\begin{align}
{\rm Re}\,&W(\varepsilon)
=\frac{\varepsilon}{2}\ln
\left[\left(\napier^{2x_1/\varepsilon}+
\napier^{2x_2/\varepsilon}\right)(1+\Delta)\right]\nonumber\\
\label{eq:sec02-mp-trans-real01}
&=\frac{\varepsilon}{2}\ln\left(\napier^{2x_1/\varepsilon}+
\napier^{2x_2/\varepsilon}\right)
+\frac{\varepsilon}{2}\ln(1+\Delta).
\end{align}
Due to the fact $x_1>x_2$ and the similar calculation to the max-plus 
transformation of real variables, the first term of eq.\ 
(\ref{eq:sec02-mp-trans-real01}) yields
$$
\frac{\varepsilon}{2}\ln\left(\napier^{2x_1/\varepsilon}+
\napier^{2x_2/\varepsilon}\right)\to x_1,
\quad\mbox{as $\varepsilon\to+0$.}
$$
We note that $\Delta$ tends to zero as $\varepsilon\to 0$, 
as long as $x_1\ne x_2$ is satisfied, and
$\ln(1+\Delta)$ is kept to be bounded.
This means that the second term of the right-hand side of
eq.\ (\ref{eq:sec02-mp-trans-real01}) vanishes as $\varepsilon\to +0$.
Hence we have
\begin{equation}\label{eq:sec02-mp-trans-real}
\lim_{\varepsilon\to +0}{\rm Re}\,W(\varepsilon)=x_1
={\rm Re}\,U_1.
\end{equation}
\par
Next, we shall consider the imaginary part of $W(\varepsilon)$, 
${\rm Im}\,W(\varepsilon)$.
Because of the relation
\begin{equation}
\begin{split}
\arg&\left(\napier^{U_1/\varepsilon}+\napier^{U_2/\varepsilon}\right)\\
&=\arctan\left[
\frac{%
\napier^{x_1/\varepsilon}\sin(y_1/\varepsilon)+
\napier^{x_2/\varepsilon}\sin(y_2/\varepsilon)
}{
\napier^{x_1/\varepsilon}\cos(y_1/\varepsilon)+
\napier^{x_2/\varepsilon}\cos(y_2/\varepsilon)
}
\right],
\end{split}
\label{eq:sec02-mp-imaginary}
\end{equation}
we have for an infinitesimal value of $\varepsilon$
and $x_1>x_2$,
\begin{equation}\label{eq:sec02-mp-trans-imag01}
\arg\left(\napier^{U_1/\varepsilon}+\napier^{U_2/\varepsilon}\right)
\simeq\arctan\tan\frac{y_1}{\varepsilon}=\frac{y_1}{\varepsilon}.
\end{equation}
Multiplying $\varepsilon$ to the both sides of 
(\ref{eq:sec02-mp-trans-imag01}) and taking the limit 
$\varepsilon\to+0$, 
we find
\begin{equation}\label{eq:sec02-mp-trans-imag02}
\varepsilon\arg\left(
\napier^{U_1/\varepsilon}+\napier^{U_2/\varepsilon}\right)
\to y_1={\rm Im}\,U_1,\quad
\mbox{as $\varepsilon\to +0$.}
\end{equation}
Hereby from eqs.\ (\ref{eq:sec02-mp-trans-real}) and
(\ref{eq:sec02-mp-trans-imag02}), we obtain
\begin{subequations}\label{eq:sec02-mp-result1}
\begin{equation}
\lim_{\varepsilon\to+0}W(\varepsilon)= U_1,\quad
\mbox{for ${\rm Re}\,U_1>{\rm Re}\,U_2$}.
\end{equation}
By similar procedures, we can see the following holds:
\begin{equation}
\lim_{\varepsilon\to+0}W(\varepsilon)= U_2,\quad
\mbox{for ${\rm Re}\,U_1<{\rm Re}\,U_2$}.
\end{equation}
\end{subequations}
\par
Lastly, let us consider the formula for the case
$x_1=x_2$. For this case, $\Delta$ defined in 
(\ref{eq:sec02-mp-trans-lnarg}) is given by
$$
\Delta=\cos\frac{y_1-y_2}{\varepsilon}.
$$
As long as $\Delta\ne-1$, that is, a condition
\begin{equation}\label{eq:sec02-mp-eqreal-limcond}
\varepsilon\ne\frac{y_1-y_2}{(2n+1)\pi},\quad\mbox{$n$: integer,}
\end{equation}
holds, we can see that $(1+\Delta)^{\varepsilon/2}$ goes to unity
under $\varepsilon\to 0$, and the second term of 
eq.\ (\ref{eq:sec02-mp-trans-real01}) vanishes in the same limit.
Then, assuming (\ref{eq:sec02-mp-eqreal-limcond})
in calculating $\varepsilon\to+0$, ${\rm Re}\,W(\varepsilon)$
yields the same result as in the case $x_1\ne x_2$.
As for ${\rm Im}\,W(\varepsilon)$, 
eq.\ (\ref{eq:sec02-mp-imaginary}) reduces to
\begin{align}
\arg&\left(\napier^{U_1/\varepsilon}+
\napier^{U_2/\varepsilon}\right)
=\arctan\left[
\frac{%
\sin(y_1/\varepsilon)+\sin(y_2/\varepsilon)
}{
\cos(y_1/\varepsilon)+\cos(y_2/\varepsilon)
}
\right]\nonumber\\
&=
\arctan\tan\frac{y_1+y_2}{2\varepsilon}=
\frac{y_1+y_2}{2\varepsilon},
\label{eq:sec02-mp-eqreal-imag}
\end{align}
and we arrive at a result
\begin{equation}\label{eq:sec02-mp-eqreal-result}
\lim_{\varepsilon\to+0}W(\varepsilon)=
\frac{U_1+U_2}{2},\quad
\mbox{for ${\rm Re}\,U_1={\rm Re}\,U_2$.}
\end{equation}
Collecting the results (\ref{eq:sec02-mp-result1}) and
(\ref{eq:sec02-mp-eqreal-result}), we obtain
\begin{equation}\label{eq:sec02-mp-results-all}
\lim_{\varepsilon\to+0}\varepsilon\log u=\begin{cases}
\hfil U_1&({\rm Re}\,U_1>{\rm Re}\,U_2),\\
\hfil U_2&({\rm Re}\,U_1<{\rm Re}\,U_2),\smallskip\\
\displaystyle{\frac{U_1+U_2}{2}}&({\rm Re}\,U_1={\rm Re}\,U_2).\\
\end{cases}
\end{equation}
This relation is considered to be a generalization of $\max$ relation
defined for real variables.
Hereafter, we shall write this operation as $\MC(U_1,U_2)$ to 
distinguish from $\max(a,b)$ for real numbers $a$ and $b$.
\section{Operation Rules of $\MC$ and Its Generalization to
More Than Two Variables}
We shall cmopare the operation rules of $\MC$ with the $\max$ operations
for real variables.
Hereafter in this section, we assume that $A$, $B$ and $C$ are complex numbers.
First, we can immediately see from the definition, the commutative law,
\begin{equation}\label{eq:sec03-mp-commutative}
\MC(A,B)=\MC(B,A),
\end{equation}
is satisfied as is the case for $\max$ operation for real numbers.
Secondly, by taking logarithm of a relation
$$
a(b+c)=ab+ac,
$$
and setting $\varepsilon\to +0$, we can see that the distributive law
\begin{equation}\label{eq:sec03-mp-distributive}
\MC(A+B,A+C)=A+\MC(B,C),
\end{equation}
is valid for arbitrary variables.
\par
However, the associative law,
\begin{subequations}\label{eq:sec03-mp-associative-all}
\begin{equation}\label{eq:sec03-mp-associative}
\MC(A,\MC(B,C))=\MC(\MC(A,B),C),
\end{equation}
is true at almost all values except at the case
\begin{equation}\label{eq:sec03-mp-associative-cond}
{\rm Re}\,A={\rm Re}\,B={\rm Re}\,C.
\end{equation}
\end{subequations}
This is because the result of taking averages usually depends on
the order of operations.
This situation leads to the difficulty to construct
the max-plus algebra for three or more complex variables.
Hereafter in this section, we are going to define this operation
on general number of variables.
\par
Let us consider a summation of $N$ variables $u_1,\ldots,u_N$:
$$
u\equiv u_1+\cdots+u_N.
$$
As in the previous section, we set the variables as
\begin{equation}
\begin{split}
u_j&=\napier^{U_j/\varepsilon},\quad U_j=x_j+\imagunit y_j,\\
j&=1,\ldots,N.
\end{split}
\end{equation}
If the number of the terms which have the largest real parts is two or less,
things are the same as the previous section, and we have
\begin{equation}\label{eq:sec03-mp-def-simple}
\begin{split}
&\MC(U_1,\ldots,U_N)\\
&\ =
\begin{cases}
\hfil U_k,\hfil&\mbox{$U_k$ has the largest real part,}\smallskip\\
\displaystyle\frac{U_k+U_l}{2},&
\mbox{$U_k$ and $U_l$ have the largest real parts.}
\end{cases}
\end{split}
\end{equation}
We assume that $U_1,\ldots,U_K$ ($K\ge 3$) have the same largest real
parts, denoted as $x$.
Calculating $\varepsilon\log u$ under an assumption that $\varepsilon$
is infinitesimal, we have
\begin{equation}\label{eq:sec03-mp-summation}
\begin{split}
\varepsilon\log&u=
\varepsilon\log\left(u_1+\cdots+u_N\right)\\
\simeq\,&\varepsilon\ln\left|
\napier^{U_1/\varepsilon}+\cdots+\napier^{U_K/\varepsilon}\right|\\
&+\imagunit\varepsilon\arg\left(
\napier^{U_1/\varepsilon}+\cdots+\napier^{U_K/\varepsilon}
\right),
\end{split}
\end{equation}
because all the terms $\napier^{U_j/\varepsilon}$ ($K+1\le j\le N$)
can be neglected due to the fact
$$
\napier^{(x_j-x)/\varepsilon}\to0,\quad\mbox{as $\varepsilon\to+0$.}
$$
\par
Similarly in the previous section, we find an approximation for
infinitesimal $\varepsilon$ as 
\begin{subequations}\label{eq:sec03-mp-real-all}
\begin{align}
{\rm Re}\,(\varepsilon\log u)&\simeq
\frac{\varepsilon}{2}\left[\ln\napier^{2x/\varepsilon}
+\ln\left|\sum_{j=1}^K\left(\cos\frac{y_j}{\varepsilon}+
\imagunit\sin\frac{y_j}{\varepsilon}\right)\right|
\right]\nonumber\\
&=x+\frac{\varepsilon}{2}\ln K+
\frac{\varepsilon}{2}\ln(1+\Delta),
\label{eq:sec03-mp-real}
\end{align}
where
\begin{equation}\label{eq:sec03-mp-real-delta}
\Delta\equiv\frac{1}{K}\sum_{\stackrel{\scriptstyle j,l=1}{j\ne l}}^K
\cos\frac{y_j-y_l}{\varepsilon}.
\end{equation}
\end{subequations}
We can see that the second term of the right-hand side of 
(\ref{eq:sec03-mp-real}) goes to zero under $\varepsilon\to0$, 
and it is also the case for the third term, if $\varepsilon$ satisfies
\begin{equation}
\frac{1}{K}\sum_{\stackrel{\scriptstyle j,l=1}{j\ne l}}^K
\cos\frac{y_j-y_l}{\varepsilon}\ne -1.
\end{equation}
As for the imaginary part of (\ref{eq:sec03-mp-summation}), 
because $U_1,\ldots,U_K$ have the same real part, we have
\begin{align*}
{\rm Im}\,&(\varepsilon\log u)\simeq
\varepsilon\arg(\napier^{U_1/\varepsilon}+\cdots+\napier^{U_K/\varepsilon})\\
&=
\varepsilon\arg(\napier^{\imagunit y_1/\varepsilon}+\cdots+
\napier^{\imagunit y_K/\varepsilon}).
\end{align*}
Therefore, we obtain a relation
$$
\varepsilon\log u\to
x+\imagunit\lim_{\varepsilon\to+0}\varepsilon\arg
\left(\napier^{\imagunit y_1/\varepsilon}+\cdots+
\napier^{\imagunit y_K/\varepsilon}\right).
$$
From the discussions above, we have derived the definition of
$\MC$ for complex variables $U_1,\ldots,U_N$.
We assume that $K$ of them,
$U_{j_1},\ldots,U_{j_K}$, have the same largest real part
whose value is $x$,
\begin{subequations}\label{eq:sec03-mp-generalcase-all}
\begin{equation}
U_{j_k}=x+\imagunit y_{j_k},\quad
k=1,\ldots,K.
\end{equation}
Then we have
\begin{align}
\MC&(U_1,\ldots,U_N)\nonumber\\&=
\begin{cases}
\hfil U_{j_1},\hfil&\mbox{for $K=1$,}\smallskip\\
\displaystyle{
x+\imagunit\lim_{\varepsilon\to+0}
\varepsilon\arg\left(\sum_{k=1}^K\napier^{\imagunit y_{j_k}/\varepsilon}
\right)},&\mbox{for $K\ge2$.}
\label{eq:sec03-mp-generalcase}
\end{cases}
\end{align}
This is the operation of $\MC$ for the general number of complex variables.
This definition is consistent with eq.\ (\ref{eq:sec03-mp-def-simple}).
Unfortunately, although we have defined the $\MC$ operation, 
the associative law (\ref{eq:sec03-mp-associative}) is not satisfied.
But if the imaginary parts of $U_j$ ($1\le j\le K$) is distributed
in equal distance, the latter case of (\ref{eq:sec03-mp-generalcase})
can be reduced to
\begin{equation}\label{eq:sec03-mp-general-equal-distribution}
\MC(U_1,\ldots,U_K)=\frac{1}{K}\left(U_1+\cdots+U_K\right).
\end{equation}
The proof of (\ref{eq:sec03-mp-general-equal-distribution}) is given in
the Appendix.
\end{subequations}
\section{%
Application of the $\MC$ operation to the Formula of DFT}
In this section, as an application, we shall apply $\MC$ operation for DFT.
We shall consider a variable $u_{k,l}$, which is a discretization
of a function $u(x,t)$, on a region.
\begin{equation}\label{eq:sec04-indep-vars-region}
-N\le k\le N,\quad
-M\le l\le M.
\end{equation}
If we apply DFT on $u_{k,l}$, we have
\begin{equation}\label{eq:sec04-dft}
u_{k,l}=\sum_{n=-N}^N\sum_{m=-M}^M c_{n,m}
\exp\left(\frac{2\pi\imagunit kn}{2N+1}+\frac{2\pi\imagunit lm}{2M+1}\right),
\end{equation}
where $c_{n,m}$ is the Fourier coefficient.
\par
Now we define 
\begin{subequations}\label{eq:sec04-parameters-setup}
\begin{equation}\label{eq:sec04-theta-def}
\theta=
\frac{2\pi kn}{2N+1}+\frac{2\pi lm}{2M+1},
\end{equation}
and $\varepsilon$ as
\begin{equation}\label{eq:sec04-vareps-def}
\varepsilon=\frac{2\pi}{2N+1}.
\end{equation}
\end{subequations}
In order to transform the phase of DFT in the complex max-plus
transformation, we introduce new coordinates, defined by
\begin{equation}\label{eq:sec04-stretched-coordinates}
\begin{split}
&\xi=\varepsilon n,\quad
\nu=\varepsilon k,\\
&\eta=\varepsilon m\sqrt{\frac{2N+1}{2M+1}},\quad
\tau=\varepsilon l\sqrt{\frac{2N+1}{2M+1}}.
\end{split}
\end{equation}
We call these variables stretched coordinates.
Substituting (\ref{eq:sec04-stretched-coordinates}) into 
(\ref{eq:sec04-theta-def}), we have
\begin{equation}\label{eq:sec04-dft-with-strcoord}
\theta=\frac{\xi\nu+\eta\tau}{\varepsilon}.
\end{equation}
When we apply the max-plus transformation to $u_{k,l}$ and $c_{n,m}$
as
\begin{equation}\label{eq:sec04-dft-newvariables}
\begin{split}
&u_{k,l}=\napier^{U_{\nu,\tau}/\varepsilon},\quad
c_{n,m}=\napier^{C_{\xi,\eta}/\varepsilon},\\
&\mbox{$U_{\nu,\tau}$, $C_{\xi,\eta}$: complex numbers},
\end{split}
\end{equation}
and setting $\varepsilon\to+0$, we can get a max-plus equation
from eq.\ (\ref{eq:sec04-dft}).
The limit $\varepsilon\to+0$ is equivalent to $N\to\infty$
under a condition where $M/N$ is finite.
We shall assume that the stretched coordinates remain to be constant
as $N\to\infty$.
As we can see from eqs.\ (\ref{eq:sec04-stretched-coordinates}), 
the stretched coordinates have discrete values numbered with integers.
Then applying suitable scale transformations for these variables,
we can assume that the values of $\xi$ and $\eta$ are restricted
to be integers.
If we express $C_{\xi,\eta}=x_{\xi,\eta}+\imagunit y_{\xi,\eta}$,
we have
\begin{equation}
\napier^{U_{\nu,\tau}/\varepsilon}=\sum_{\xi,\eta\in{\mathbb Z}}
\napier^{x_{\xi,\eta}+i(y_{\xi,\eta}+\xi\nu+\eta\tau)}.
\end{equation}
Applying the max-plus transformation for complex variables
defined in the previous sections, we obtain
\begin{equation}\label{eq:sec04-dft-result}
\begin{split}
&U_{\nu,\tau}=\MC\left(\{
x_{\xi,\eta}+i\phi_{\xi,\eta,\nu,\tau}\}\right),\\
&\phi_{\xi,\eta,\nu,\tau}=y_{\xi,\eta}+\xi\nu+\eta\tau.
\end{split}
\end{equation}
Equation (\ref{eq:sec04-dft-result}) means that the application
of max-plus transformation to DFT yields the complex number
which have the largest amplitude of discrete Fourier coefficient.
We note that the trajectory on $\nu\tau$-plane, 
$U_{\nu,\tau}=\mbox{const.}$, 
is presented by the line $\phi_{\xi,\eta,\nu,\tau}=\mbox{const.}$,
for prescribed $\xi$ and $\eta$.
\section{Concluding Remarks and Discussions}
In this paper, we have proposed a method to construct a max-plus equation
for complex variables.
By calculating logarithm of variables explicitly for complex variables,
a novel operation $\MC$ of max-plus type has been introduced.
The conventional max-plus transformation has a difficulty that
we cannot apply max-plus transformation
(\ref{eq:sec01-mp-trans-all}) to non-positive definite variables
and subtraction terms.
For example, an ultradiscretization of the sine-Gordon equation
has tried by applying max-plus transformation to a discrete
analogue of sine-Gordon equation to avoid this problem\cite{IMNS}.
But it seems that this result owes much to the good structure
of the original equation.
We are convinced that the operation given in this paper dissolves
such difficulty and enables us to generate discrete systems from
difference equations automatically.
\par
Recently, a way of generalization of $\max$ algebra which allows
subtraction operation on general field was introduced\cite{Ochiai}.
The relation between the results listed in this reference and the one
presented in this paper is not clear, 
but the comparison of these methods is one of the worthwhile problems.
\par
Furthermore, we have applied $\MC$ to DFT.
Introducing stretched coordinates clarified the relation between
DFT and complex max-plus transformation.
Since discrete Fourier transformation is a common method to solve
differential equations,
we can expect that solutions of various difference equations
are easily discretized by virtue of this method.
Moreover, we can expect that $\MC$ operation enables us to discretize
other types of integral transformation, such as Laplace transformation,
and this will yield a systematic way to solve full discrete systems.
\par
At the end of this paper, we shall point out several problems
concerning the results.
The definition of the operation $\MC$ for general number of variables
(\ref{eq:sec03-mp-generalcase-all}) does not allow assosiative law,
except for the special case shown in eq.\ 
(\ref{eq:sec03-mp-general-equal-distribution}).
This relation still includes limiting process,
and it seems to be difficult to derive the result of transformation
for variables given arbitrary.
The construction of algebras which satisfy accociative law
is a future problem.
\section*{Acknowledgments}
The authors are grateful to Professor Hideo Nakajima for his continual
encouragements and stimulated discussions.
Thanks are also due to Professor Daisuke Takahashi and Professor Katsuhiro
Nishinari for their kind interests in this work, critical reading of
the manuscript and valuable comments.
This work is partially supported by Grant-in-Aid for Scientific Research (C)
$13640395$ from the Ministry of Education, Culture, Soprts, Science and 
Technology.
\appendix
\sectiona{}
We shall prove the formula
\begin{equation}
\MC(U_1,\ldots,U_K)=\frac{U_1+\cdots+U_K}{K},
\end{equation}
under the condition where all the $U_j$'s have the same real part
and their imaginary parts are distributed in equal distance.
Let us set
\begin{subequations}\label{eq:app-variable-distribution}
\begin{equation}
U_j=x+\imagunit y_j,\quad j=1,\ldots,K,
\end{equation}
where
\begin{equation}\label{eq:app-imaginary-vars}
y_j=\alpha+\beta(j-1).
\end{equation}
\end{subequations}
The summation of (\ref{eq:sec03-mp-generalcase}) is
that of a geometric series with common ratio
$\napier^{\imagunit\beta}$. Then we have
\begin{align}
\sum_{j=1}^K&\napier^{\imagunit y_j/\varepsilon}
=\napier^{\imagunit\alpha/\varepsilon}
\sum_{j=1}^K\napier^{\imagunit(j-1)\beta/\varepsilon}
=\frac{1-\napier^{\imagunit\beta K/\varepsilon}}
{1-\napier^{\imagunit\beta/\varepsilon}}\nonumber\\
&=\frac{\sin(\beta K/2\varepsilon)}{\sin(\beta/2\varepsilon)}
\exp\left\{\frac{\imagunit}{\varepsilon}
\left[\alpha+\frac{\beta(K-1)}{2}\right]\right\}.
\end{align}
The argument of this quantity is derived as
\begin{equation}\label{eq:app-argument}
\arg\left(\sum_{j=1}^K\napier^{\imagunit y_j/\varepsilon}\right)
=\frac{1}{\varepsilon}
\left[\alpha+\frac{\beta(K-1)}{2}\right].
\end{equation}
Under the condition (\ref{eq:app-variable-distribution}), 
we find from eq.\ (\ref{eq:app-argument}) that
\begin{equation}
\MC(U_1,\ldots,U_K)=x+\imagunit\left[\alpha+\frac{\beta(K-1)}{2}\right].
\end{equation}
This is no other than the arithmetic mean of
$U_1,\ldots,U_K$ defined in (\ref{eq:app-variable-distribution}).
Hereby we have proved eq.\ (\ref{eq:sec03-mp-general-equal-distribution}).

\end{document}